\documentclass{paper}

\usepackage[utf8]{inputenc}
\usepackage{todonotes}
\usepackage{listings}
\usepackage{graphicx}
\usepackage{cprotect}
\usepackage{xcolor}
\usepackage{relsize}
\usepackage{xspace}
\usepackage[frozencache]{minted}
\usepackage[maxnames=3,backend=biber,citestyle=numeric-comp,bibstyle=ieee,sorting=none]{biblatex}
\usepackage[a4paper, top=2.5cm,bottom=2.5cm,left=2.6cm,right=2.6cm,marginparwidth=2cm]{geometry}
\usepackage{booktabs} 
\usepackage[title, header]{appendix}
\usepackage{textcomp} 
\usepackage{subfig} 
\usepackage{siunitx}
\usepackage{xurl} 
\usepackage[hidelinks]{hyperref}

\usepackage{caption} 
\usepackage[affil-it]{authblk}

\newcommand{\petrifyml}{petrifyML\xspace}
\newcommand{\petrifyML}{\petrifyml}
\newcommand{\petrifybdt}{\petrifyml}

\newcommand{\atlas}{ATLAS\xspace}
\newcommand{\cms}{CMS\xspace}
\newcommand{\lhc}{LHC\xspace}

\newcommand{\gambit}{\textsc{GamBit}\xspace}
\newcommand{\colliderbit}{\textsc{ColliderBit}\xspace}
\newcommand{\rivet}{\textsc{Rivet}\xspace}
\newcommand{\checkmate}{CheckMATE2\xspace}
\newcommand{\hepdata}{HEPData\xspace}
\newcommand{\madgraph}{\textsc{MadGraph\_aMC@NLO}\xspace}
\newcommand{\protos}{\textsc{Protos}\xspace}

\newcommand{\cpp}{C\raisebox{0.35ex}{\smaller ++}\xspace} 
\newcommand{\python}{\textsc{Python}\xspace}
\renewcommand{\root}{ROOT}\xspace
\newcommand{\AlmaLinux}{\texttt{AlmaLinux}}

\newcommand{\emlearn}{\textsc{emlearn}\xspace}
\newcommand{\onnxToC}{\textsc{onnx2c}\xspace}
\newcommand{\hlsForML}{\textsc{hls4ml}\xspace}
\newcommand{\keras}{\textsc{Keras}\xspace}
\newcommand{\lgbm}{\texttt{lgbm}\xspace}
\newcommand{\lwtnn}{\texttt{lwtnn}\xspace}
\newcommand{\mvautils}{\texttt{MVAUtils}\xspace}
\newcommand{\onnx}{ONNX\xspace}
\newcommand{\onnxruntime}{OnnxRuntime\xspace}
\newcommand{\scikitlearn}{\textsc{scikit-learn}\xspace}
\newcommand{\sklearn}{\scikitlearn}
\newcommand{\sofie}{SOFIE\xspace}
\newcommand{\tensorflow}{TensorFlow\xspace}
\newcommand{\tfonnx}{\texttt{tf2onnx}\xspace}
\newcommand{\tmva}{\textsc{TMVA}\xspace}
\newcommand{\xgboost}{\texttt{xgboost}\xspace}

\newcommand{\numpy}{\texttt{numpy}\xspace}
\newcommand{\pandas}{\texttt{pandas}\xspace}
\newcommand{\pip}{\texttt{pip}\xspace}
\newcommand{\pytest}{\texttt{pytest}\xspace}
\newcommand{\pytestCov}{\texttt{pytest-cov}\xspace}
\newcommand{\tfkeras}{\texttt{tf\_keras}\xspace}
\newcommand{\uproot}{\texttt{uproot}\xspace}
\newcommand{\pyyaml}{\texttt{pyyaml}\xspace}

\newcommand{\json}{\texttt{json}\xspace}
\newcommand{\opset}{\texttt{opset}\xspace}

\newcommand{\installWith}[1]{\textit{Install with} \texttt{pip install petrifyml[#1]}}
\newcommand{\eg}{e.g.\,}

\DeclareSourcemap{
	\maps[datatype=bibtex,overwrite=true]{
		\map{
			\step[fieldsource=Collaboration, final=true]
			\step[fieldset=usera, origfieldval, final=true]
		}
	}
}

\renewbibmacro*{author}{%
	\iffieldundef{usera}{%
		\printnames{author}%
	}{%
		\textsc{\printfield{usera}} Collaboration%
	}%
}

\title{Enabling stable preservation of ML algorithms\\ in high-energy physics with \petrifyml}
\author{%
    A.~Buckley$^{\,1}$,
    L.~Corpe$^{\,2}$,
    M.~Habedank$^{\,1}$,
    T.~Procter$^{\,3}$
}

\begin{document}

\maketitle{}

\begin{center}
  \itshape
  $^1$ School of Physics \& Astronomy, University of Glasgow, G12~8QQ, Glasgow, UK\\
  $^2$ Universit\'e Clermont Auvergne, Laboratoire de Physique de Clermont Auvergne,\\
  CNRS/IN2P3, Clermont-Ferrand, France\\
  $^3$ Jagiellonian University, \L{}ojasiewicza 11, 30-348 Krak\'ow; Poland.
\end{center}

\begin{abstract}
    Machine learning (ML) in high-energy physics (HEP) has moved in the \lhc\ era from an internal detail of experiment software, to an unavoidable public component of many physics data analyses. Scientific reproducibility thus requires that it be possible to accurately and stably preserve the behaviours of these, sometimes very complex algorithms. We present and document the \petrifyml\ package, which provides missing mechanisms to convert configurations from commonly used HEP ML tools to either the industry-standard \onnx\ format or to native \python\ or \cpp\ code, enabling future re-use and re-interpretation of many ML-based experimental studies.
\end{abstract}


\section{Introduction}

Despite being early adopters of machine learning (ML) since the 1980s~\cite{Quinlan:1986gsq,Corneliusen:1989zw},
high-energy physics~(HEP) experiments until recently typically limited its use to ``internal'' mechanisms such as physics-object calibration and reconstruction rather than directly publishing ML outputs as scientific results. The classic example of this is the flavour ``tagging'' of hadronic jets~\cite{ATLAS:2019bwq,CMS:2020poo}, which has long been ML-based, and more recently other complex reconstructions such as electron/photon or tau/hadronic jet disambiguation where ML methods can improve performance over closed-form calibrations. In this mode, ML effects could largely be summarised as simple tabulations of object reconstruction efficiencies or resolutions, so re-interpretation of HEP experimental publications did not require re-running the original trained ML algorithm.

The Run~2 phase of the Large Hadron Collider has changed this picture considerably, coinciding as it has with the rapid rise of both well-aligned computing architectures and new, highly effective ML architectures. These effects have been transformational in HEP: not only have modern architectures and computing power produced step-changes in the performance of tagging and other algorithms, but they have also underpinned the development of data-analysis techniques which provide non-parametrically optimal sensitivity to proposed models of new physics. While excellent news for the ability of experiments to study given (and typically rather simplified) models, this foregrounding of ML methods as an integral part of the physics study itself -- not replaceable with an efficiency tabulation -- introduces technical challenges for scientific reproducibility, particularly in the long term, which have only recently begun to receive significant attention~\cite{Araz:2023mda}.

These problems centre on toolkit dependencies and format stability. Current ML algorithms are typically trained, and often deployed, from \python-based tools, but these are then not easily usable from other computing languages. Even for those who wish to re-use from \python, the variety of toolkits is large and a particular preserved ML algorithm may behave differently even in different versions of its toolkit. This, and the desirability of transfer between different training and runtime environments, motivated the development and evolution of the \onnx{} format~\cite{onnx_github}. However, several ML frameworks -- in particular tools developed specifically within HEP, such as \tmva~\cite{Speckmayer:2010zz} and \lwtnn~\cite{daniel_hay_guest_2024_14276439} -- do not natively support conversion to \onnx. Concerns remain about the long-term reproducibility of \onnx\ behaviours as their execution environments evolve, without cumbersome and slow approaches such as wholly containerising each trained ML's runtime environment: this motivates also, in some cases, preserving algorithms as dependency-free native code.

The \petrifyml\ tool~\cite{petrifyml} described in this release note was created to fill gaps in this chain of algorithm preservation. A PyPI-installable \python\ package and set of command-line tools, it provides converter routines to \onnx\ for neural networks (NNs) created with the \lwtnn\ and \tmva\ HEP frameworks, and to \onnx\ and native \cpp\ and \python\ for boosted decision trees (BDTs) created with \tmva\ and \sklearn.
To allow the use of these preserved ML models \textit{in practice}, additional information is required.
\petrifyML\ retains as much ML model metadata, \eg training settings and normalisation, as possible but it is limited by the capabilities of the input and output formats.

The closest tool that provides some similar functionality is \emlearn~\cite{emlearn}\footnote{Though notably this post-dates the original release of what was then known as \texttt{petrifyBDT}.}, which converts some basic ML formats (most pertinently, those that overlap are \scikitlearn\ decision tress and \keras~\cite{chollet2015keras} sequential neural networks) to native \textsc{C} code, for the purposes of inference with micro-electronics. However, because of its micro-electronics focus, it does not produce \python\ code, and does not support any of the HEP-specific formats. It is also worth noting some tools similar to \emlearn\ such as \onnxToC~\cite{onnx2c} and \hlsForML~\cite{fastml_hls4ml, Duarte:2018ite}, which are focussed on microcontrollers, FPGAs, and so forth: however these do not overlap with \petrifyML{}.

This tool is not intended as a replacement for native export methods when all the necessary information is available: for example, if a user has trained a neural network in \keras, we would advise them to output directly to an \onnx\ file, rather than using a combination of \lwtnn{} and \petrifyml{} to try to achieve the same goal.

Users are furthermore strongly encouraged to provide extensive documentation and validation data as outlined in the Les Houches guidelines on re-interpretable ML~\cite{Araz:2023mda} when publishing preserved ML models: just providing a set of weights produced by \petrifyml{} will not be sufficient!
% ...

\section{Installation and dependencies}

\petrifyml\ is fully written in \python\ and strives to depend only on a minimal suite of well-maintained, standard, and lightweight packages.
Naturally, these encompass \onnx\ and \onnxruntime~\cite{onnxruntime}, with \numpy~\cite{harris2020array} and \pandas~\cite{reback2020pandas} used for utilities.
\tensorflow~\cite{tensorflow2015-whitepaper}, \tfkeras~\cite{chollet2015keras}, and \tfonnx~\cite{tf2onnx} are used for the conversion of TMVA multilayer perceptrons via \keras.
\uproot~\cite{uproot2020} is used for converting \mvautils\ BDTs, and the heavyweight \root\ framework~\cite{Brun:1997pa} is only required for conversions from TMVA BDTs to plain-text \cpp{} or \python{}. \scikitlearn{}~\cite{scikit-learn} is required only for the conversion of \scikitlearn{} decision trees, which also requires the \texttt{joblib} module~\cite{The_joblib_developers_joblib}.
Lastly, \pytest~\cite{pytest}, \pytestCov, and \pyyaml\ are employed for the automated testing of the conversion routines.

All \petrifyML\ scripts optionally output an example validation script or similar.
Unavoidably, these validation codes have a larger set of dependencies. For example, converting an \mvautils{} \root{} file to \onnx\ requires neither \root{} nor \onnxruntime{}, but the validation macros require both \root{} (with an \mvautils\ install) and \onnxruntime{}.

\noindent The \petrifyml\ package (and its dependencies) can be conveniently installed using \pip:
\begin{minted}{bash}
	$ pip install petrifyml[part]
\end{minted}
Specifying a \texttt{[part]} allows to have a lightweight installation with only the required packages for a single operation mode of \petrifyml: ``\texttt{sklbdt}'', ``\texttt{tmvabdt}'', ``\texttt{lwtnn}", ``\texttt{tmvamlp}", ``\texttt{mvautils}", or ``\texttt{dev}" for all parts and the test suite.

\section{Converting BDTs to native \cpp{} and \python{}}
\label{sec:cppAndPython}

Conversion to native \cpp\ and \python\ code avoids any sort of dependency, guaranteeing verbatim performance in perpetuity for smaller BDTs.
This is particularly advantageous as the industry focus driving the development of the \onnx\ format is on short-timescale shifts between platforms rather than long-term preservation.
This means that the long-term reproducibility of \onnx\ behaviours as their execution environments evolve need to be monitored.
Conversely, native, human-readable \cpp\ and \python\ code grow quickly to a considerable size, limiting this conversion to small and medium-sized BDTs for practical use.

\subsection{petrify-sklbdt-to-cpp}

\installWith{sklbdt}

\subsubsection{Background}

The \scikitlearn\ \python{} package provides a variety of models for decision trees and forests.
The script described in this section is designed for the \texttt{GradientBoostingClassifier}, though should be readily extensible to other formats if the need arises. \scikitlearn{} classifiers are normally saved in \python{}'s pickle format. Given that this format can prove highly version dependent and unstable over even short periods of time, saving a pickle file (absent an entire container with the right dependencies installed) cannot be considered model preservation. Therefore, there is a clear need for a reliable, long-term preservation option for these BDTs.

\subsubsection{Running}
The \petrifyML\ script accepts trees in either \texttt{.pkl}\footnote{Note that \python's pickle format is generally not backward compatible; ensure to save using a pickle protocol version supported by the \python version used to run \petrifyml.} or \texttt{.job} formats. For a classifier stored in a \texttt{.pkl} file, a simple use case would be
\begin{minted}{bash}
	$ petrify-sklbdt-to-cpp mytree.pkl -n my-cpp-file 
\end{minted}
Note that the \texttt{.cc} suffix is appended automatically to the \texttt{-n} argument. If it is a valid \cpp{} function name (after dashes and spaces have been converted to underscores), the \texttt{-n} argument (or its basename, if a path) is also used as the function name inside the petrified \cpp{} file: otherwise, the function name defaults to \texttt{decision}.

Further information on all optional arguments can be accessed via the \texttt{-h} option; most are self-explanatory, but some deserve some additional notes:
\begin{itemize}
	\item \texttt{--write-validation} (default: \texttt{False}): Appends a \texttt{main} function to the output \texttt{.cc} file that performs test calls of the BDT function on randomly generated input, and prints compilation instructions, in order to allow easier validation.

    \item \texttt{--run-validation} (default: \texttt{False}): As above, but also compiles the new \texttt{.cc} file, runs it, and compares output to the original \scikitlearn{} model. Assumes that either \texttt{g++}, \texttt{c++} or \texttt{clang++} commands are available and provide a working \cpp{} compiler\footnote{If users prefer to use a different compiler, then the \texttt{\$CXX} environment variable can be pointed to the compiler-of-choice.}. Optionally, if a list of numbers is provided as the argument, these are used as the input to the BDT (in this case it is the user's responsibility to ensure they provide the correct number of inputs). If no inputs are provided, then the test inputs are drawn from a normal distribution constructed from the mean and standard deviations of the cut-values in the BDT.
\end{itemize}

\subsubsection{Output}
\label{SUBSUBSEC_SKL_OUTPUT}

The final BDT function is found at the bottom of the outputted \cpp{} file, with a name either matching the \texttt{-n} argument, or defaulting to \texttt{decision} if the \texttt{-n} argument is not a valid \cpp{} function name. It expects a vector of the BDT inputs, returns an output, and is templated to allow the use of different input types. Users are free to include this inside their project as they wish: for example, it may make sense to rename the file with a header suffix, and \texttt{include} it where needed in the project; it may make sense to compile it into a dynamic library\footnote{In which case users will also need to either add their own header file or at least provide a forward declaration.}, or for smaller projects, it may even make sense to simply copy the entire file into an existing \texttt{.cc} file.

The functions above the BDT function (usually with a suffix of the form \texttt{\_XXX\_YYY}) are used internally by the BDT function. Note that if a \texttt{main} function was added for testing (with either \texttt{--run-validation} or \texttt{--write-validation}), it should be removed for real-world use.

\subsection{petrify-tmvabdt-to-cpp and petrify-tmvabdt-to-py}

\installWith{tmvabdt}


\subsubsection{Background}

The Toolkit for Multivariate Data Analysis (\tmva) is the \root\ library that provides interfaces and implementations for ML models. This includes a variety of BDTs (including multi-class classifiers), which can be stored as either XML\ or \root\ files. These formats are unfortunately tied to the heavyweight ROOT package, and therefore there is a motivation to be able to provide a lightweight alternative for simple trees. This comes in the form of converters to native \cpp{} or \python{}.
All \tmva\ BDT architectures, i.e.\,single- and multi-class classification as well as regression, are supported.
For the boosting method, \texttt{AdaBoost} as well as \texttt{Grad} are supported.

\subsubsection{Running}
For a classifier stored in an XML file (the converter does not take \root{} files), a simple use case would be 
\begin{minted}{bash}
	$ petrify-tmvabdt-to-cpp mytree.xml -n my-cpp-name
\end{minted}
or 
\begin{minted}{bash}
	$ petrify-tmvabdt-to-py mytree.xml -n my-py-name
\end{minted}
Note that the appropriate \texttt{.cc} or \texttt{.py} suffix is appended automatically to the \texttt{-n} argument. If it is a valid function name in the language of choice (after dashes and spaces have been converted to underscores), the \texttt{-n} argument (or its basename, if a path) is also used as the function name inside the petrified \cpp{} or \python{} code: otherwise, the function name defaults to \texttt{decision}.

Further information on all optional arguments can be accessed via the \texttt{-h} option; most are self-explanatory, but some deserve some additional notes:
\begin{itemize}
	\item \texttt{--write-validation} (default: \texttt{False}): Appends a \texttt{main} function to the output \texttt{.cc} or \texttt{.py} file that performs test calls of the BDT function on randomly generated input. Prints compilation instructions in the \cpp{} case, in order to allow easier validation.

    \item \texttt{--run-validation} (default: \texttt{False}): As above, but also runs the new \texttt{main} function and compares output to the original TMVA version. In the \cpp{} case, this option assumes that either the \texttt{g++}, \texttt{c++} or \texttt{clang++} commands are available and provide a working \cpp{} compiler\footnote{As before, if users prefer to use a different compiler, then the \texttt{\$CXX} environment variable can be pointed to the compiler-of-choice.}.  Optionally, if a list of numbers is provided as the argument, these are used as the input to the BDT (in this case it is the user's responsibility to ensure they provide the correct number of inputs). If no inputs are provided, then the test inputs are drawn from a normal distribution constructed from the mean and standard deviations of the cut-values in the BDT.
\end{itemize}

\subsubsection{Output format}
\label{SUBSUBSEC_TMVABDT_OUTPUT}

The \cpp{} output format for \tmva{} BDTs is effectively the same as for \sklearn{} BDTs outlined in Section~\ref{SUBSUBSEC_SKL_OUTPUT}, with the addition of multi-class BDTs, which return a vector instead of a single floating-point type.
The \python\ code format is very similar, with a function taking its name from the \texttt{-n} argument at the bottom of the file, which can be \texttt{import}ed or copied into code as desired.

\section{Converting BDTs to ONNX}

\onnx{} files are increasingly the universal standard for ML preservation, come with the blessing of the Les Houches guidelines on re-interpretable ML~\cite{Araz:2023mda}, and are already supported by most major re-interpretation frameworks. Though in principle multiple programs can be used to carry out inference using the information in an \onnx{} file, in practice the most commonly used (including in the \rivet~\cite{Bierlich:2019rhm,Bierlich:2024vqo}, \colliderbit~\cite{GAMBIT:2017qxg} and \checkmate~\cite{Dercks:2016npn} \onnx{} interfaces) is \onnxruntime.

\subsection{petrify-mvautilsxgboost-to-onnx and petrify-mvautilslgbm-to-onnx}

\installWith{mvautils}

\subsubsection{Background}

\mvautils~\cite{ATLAS:2019ath} is a package developed by the \atlas\ collaboration for the preservation and portability of BDTs.
Notably, \mvautils\ does not train BDTs itself: rather it takes BDTs that have been trained and stored in custom formats by diverse tools (\xgboost~\cite{Chen:2016:XST:2939672.2939785}, \lgbm~\cite{NIPS2017_6449f44a} and \tmva), and converts them into a \root-based format (which does differ based on the original training environment) which is typically significantly more compressed.

The objectives of the \mvautils\ package share some overlap with those of this package, and readers may ask why, if the work has already been carried out to support conversion to a format that can be read by one universal package, is another project needed?
Indeed, the \checkmate re-interpretation tool already uses parts of the \mvautils\ code (released as part of the Athena~\cite{ATLAS:2019ath} framework) to run BDTs published by \atlas.

However, the dependence on \root\ is an insurmountable issue for many other tools, such as \rivet and \gambit~\cite{GAMBIT:2017yxo}, for full production level scans. \root\ significantly complicates compilation, can cause issues with multithreading, and may be impractical for users to run locally on their own machines.
Therefore, the separate converter presented in this release note addresses a significant practical issue.

As BDT forests get larger, writing them in plain-text code files as in Section~\ref{sec:cppAndPython} may become impractical. For example, the \mvautils\ forests used by \atlas{} in the higgsino search in References~\cite{ATLAS:2024tqe, hepdata.136030.v1} include 125,000 trees, each with 120-130 nodes. At a conservative estimate, this would require a \cpp{} file over 30~million lines long, which is impractically large for a nominally ``lightweight'' format. 

The alternative, implemented here, is to convert the trees from various proprietary or ``public-yet-impractically-difficult-to-reuse'' (sadly common in HEP) formats into \onnx{} files.

At the moment, \petrifybdt{} only provides converters for \mvautils\ files created from \xgboost{} and \lgbm{} forests. The main reason for this is need: \atlas{} has not (yet) published any \mvautils\ \root\ files that originated in \tmva\ format. The converter uses the \pip-installable \uproot\ package to parse the input files (a \root{} installation is \emph{not} required to run this converter\footnote{Although if users wish to rerun \mvautils\ for comparative validation, a full installation of ROOT is unavoidable.}), and directly constructs the BDT using the \onnx{} TreeEnsemble operator without intermediate steps.

If the original \lgbm{} or \xgboost{} trees are available, then it is of course preferable to convert to \onnx{} using the inbuilt methods of those packages: we do not intend for this package to become part of a baroque chain of conversions. However, it is regrettably common for the original files to be either not publicly released or indeed be completely lost.

\subsubsection{Running}

A simple use-case for both input formats is shown below:
\begin{minted}{bash}
    $ petrify-mvautilslgbm-to-onnx myfile.root -n my-onnx-file 
\end{minted}
\begin{minted}{bash}
    $ petrify-mvautilsxgboost-to-onnx myfile.root -n my-onnx-file 
\end{minted}
Note that the \texttt{.onnx} suffix is automatically appended to the \texttt{-n} argument.

Further information on all optional arguments can be accessed via the \texttt{-h} option; most are self-explanatory, but some deserve some additional notes:
\begin{itemize}
	\item \texttt{--nf}, \texttt{--nfeatures} (default: \texttt{None}): The number of input features for the forest. If it is not provided, the script will try and guess based on the first few trees: however the nature of the \mvautils\ format means that this guess can be wrong, particularly if there is a feature that is not used in the first few trees of the forest\footnote{Indeed, for one of the forests released on \hepdata~\cite{hepdata.156776.v1} alongside Reference~\cite{ATLAS:2024woy}, the final input feature was \textit{never} used and hence totally invisible in the \mvautils\ format.}. If the guess is wrong, then this will not be apparent until inference fails. SimpleAnalysis~\cite{ATLAS:2022yru} or similar pseudocode may be the best source of this information.
	
	\item \texttt{--classifier} (default: \texttt{False}): When BDTs are used as classifiers, it is customary to apply a sigmoid function to convert the BDT-score into a (pseudo-)probability. In the \mvautils\ format, this information is not included in the \root\ file, but rather is dependent on whether the BDT is called with \texttt{GetClassification} or \texttt{GetResponse}. While the user is free to replicate this themselves, if the network is only ever to be used as a classifier, it is simpler if this function is handled within \onnxruntime. Therefore, running with \texttt{--classifier} applies a  final sigmoid function within the \onnx{} description of the model.
	
	\item \texttt{--write-validation} (default: \texttt{False}): See Section~\ref{SUBSUBSEC_MVAUTILSVALIDATION}.
	
	\item \texttt{--opset} (default: \texttt{None}) and \texttt{--ir-version} (default: \texttt{None}): See section \ref{SUBSUBSEC_MVAUTILSVERSIONS}.
\end{itemize}

\subsubsection{Validation}
\label{SUBSUBSEC_MVAUTILSVALIDATION}

The \texttt{--write-validation} option will, after the successful conversion of the network, write a directory containing four files:

\begin{itemize}
	\item \texttt{testonnx.py}: A \python\ script that runs the \onnx{} implementation of the BDT on a set of inputs, with the result printed to screen and saved to \texttt{testonnx.csv}. It is the user's responsibility to make sure the \onnxruntime \python\ module (not needed for the conversion itself) is available.
	\item \texttt{testmvautils.cxx}: A \root{} macro that runs the original \mvautils{} implementation of the network, on the same set of inputs as \texttt{testonnx.py}, also printing the result to screen and also saving it to \texttt{testmvautils.csv}. Access to \mvautils{} is the user's responsibility (e.g. via lxplus, the public machines provided by the CERN IT Department for interactive work, or an Athena docker image).
	\item \texttt{validator.py}: A minimal \python\ script that checks that the two \texttt{.csv} files agree.
	\item \texttt{readme.md}: Brief documentation reminding users how to use the scripts in the validation directory.
\end{itemize}

This is not a comprehensive check, but will likely catch most issues which are not pathological; and can serve as a starting point for further validation. In an attempt to provide inputs at a physically relevant scale, the test-inputs used for validation are drawn from normal distributions built from the mean and variance of the cut values in the BDT. Users may nevertheless replace the randomly generated inputs in \texttt{testonnx.py} and \texttt{testmvautils.cxx} with their own values if they prefer.

\subsubsection{Version-compatibility options}
\label{SUBSUBSEC_MVAUTILSVERSIONS}

The \petrifybdt{} implementation of boosted decision trees uses the \texttt{TreeEnsemble} operator from the \texttt{ai.onnx.ml} domain. This operator was only introduced in version 1.16 of \onnx{} (March 2024), and supported in \onnxruntime\ from version 1.18.0 (May 2024). Therefore, for inference to work, it is important to be using a (relatively) up-to-date version of \onnxruntime. While this is unlikely to be problematic in industry, the adoption of up-to-date package versions has not always been straightforward in HEP.

As these features are still somewhat fresh in \onnx{} and \onnxruntime, forests produced using cutting edge \onnx{} may not always be interpretable by even only slightly older (but possibly still $\geq$ 1.18.0) versions of \onnxruntime.
For example, running \petrifybdt{} with its default settings using \onnx~1.18 produces \onnx{} files that cannot be loaded by \onnxruntime\ 1.22.0\footnote{Notably, for a period of time these were the latest releases of either package.}, due to mismatches in the default \onnx\ \opset\ and intermediate-representation versions.

Therefore, we allow users to specify both the opset and intermediate-representation version via the \texttt{--opset} and \texttt{--ir-version} options\footnote{If unsure which versions are compatible with your installation, \opset\ 22 and \texttt{ir-version} 10 should be a reliable base.}.
There are no backward compatibility issues, so trees produced with older \onnx\ \opset{}s should remain usable in all future releases of \onnxruntime. Not specifying either variable (or explicitly setting to \texttt{None}) defaults to the most recent versions supported by the installation of \onnx\ that \petrifyml{} uses.

\section{Converting neural networks to \onnx}

All the same arguments that apply to preserving BDTs in \onnx\ apply to preserving neural networks in \onnx.
If anything, \onnx\ is an even stronger industry standard with respect to neural networks.

\subsection{petrify-tmvamlp-to-onnx}

\installWith{tmvamlp}

\subsubsection{Background}

\tmva, \root's library for interfaces to and implementations for ML, in addition to the aforementioned BDTs provides the option to train multilayer perceptrons (MLPs).
Their weights are stored in \root\ or XML\ files but are unfortunately tied to the heavyweight ROOT package.
With the System for Optimized Fast Inference code Emit (\sofie), \tmva\ was extended to allow reading \onnx\ files produced in other ML libraries.
Notably, \sofie\ does \textit{not} allow \tmva\ to \textit{write} \onnx\ files such that they could be used in other ML libraries.

This shortcoming is bridged by \petrifybdt.
Technically, \petrifybdt\ reads the XML\ file, creates an MLP in \tensorflow\ using \keras\ with the same architecture and node weights, and exports it to \onnx\ files using \tfonnx.
Classification as well as regression are supported.

\subsubsection{Running}
A simple use-case is:
\begin{minted}{bash}
  $ petrify-tmvamlp-to-onnx my-tmvamlp-file.xml -n my-onnx-file 
\end{minted}
Further information on all optional arguments can be accessed via the \texttt{-h} option; most are self-explanatory, though some deserve additional notes:
\begin{itemize}
	\item \texttt{--run-validation} (default: \texttt{False}).
    Validates that using the original XML file with TMVA, the generated \keras\ model, and the generated \onnx\ file with the \onnxruntime\ yield the same inference results on test data.
    The test data can be given with \texttt{--test-data}.
    If not given, it is random-uniformly generated within the parameter ranges allowed for each input variable.
    \item \texttt{--validation-only} (default: \texttt{False}).
    Only runs the validation, i.e. does not generate a \keras\ model or output an \onnx\ file.
    Allows to run the validation quickly and repeatedly.
\end{itemize}

\subsection{petrify-lightweightnn-to-onnx}

\installWith{lwtnn}

\subsubsection{Background}
\texttt{lwtnn} is a package designed for fast and efficient NN inference in \cpp\ environments with minimal dependencies, even when the original network was trained in an entirely \python{}ic environment, for example using \keras.
It was originally designed for use in the \atlas\ trigger, but particularly during \lhc\ Run 2 and before the usage of \onnx\ became more widespread, it became a common way for \atlas\ analyses and tagging groups to run neural network inference inside \cpp\ analysis frameworks.

In many ways, this is a very good format for re-interpretation: it has minimal dependencies, uses human-readable files and is easy to link from other \cpp\ projects. Indeed, \rivet{} had a header-only interface to \lwtnn{} between versions 3.1.7 and 4.1.2, which allowed individual analysis plugins to be compiled against \lwtnn{} as required\footnote{This feature was deprecated in \rivet{} 4.1.2 as \lwtnn{} usage became increasingly rare.}. Nevertheless as \lwtnn{} usage wanes even within \atlas{} -- due to a combination of the emergence of \onnx{} and \lwtnn{}'s lack of support for cutting-edge architectures -- it will be hard for all re-interpretation tools to justify linking against an aging package that may only be required for one or two analyses: thus motivating a method of converting to \onnx{}. This will also allow models previously only published in \lwtnn{} form to be used in \python\ for the first time.

\lwtnn{} stores neural network weights in a \json\ format. The exact \json\ schema depends on the network architecture: \texttt{lightweightneuralnetwork} for simpler linear models; and \texttt{lightweightgraph}, for more complex models (including BDTs). At the moment, only conversion of the former is supported in \petrifybdt{}. Similarly, as summarised in Table~\ref{TAB_LWTNN}, only a subset of the layer-types and activation functions that \texttt{lightweightneuralnetwork} supports are currently supported for conversion to \onnx. Nevertheless, the present coverage is sufficient to cover all \lwtnn{} \json\ files so far publicly released by \atlas\footnote{At the time of writing, this comes to a grand total of one neural network~\cite{hepdata.134010.v1} from an \atlas\ search for vector-like quarks~\cite{ATLAS:2022hnn}.}.
Hopefully, coverage will be extended with time, especially if the need emerges due to \atlas{} publishing more \lwtnn{} files.
The \petrifyml code is also sufficiently modular though that a moderately technically experienced user should be able to add their own layers and activation functions if required\footnote{Merge requests will always be welcome!}.

\begin{table}[hbt]
    \cprotect\caption{\lwtnn{} layer types and activation functions currently supported when converting \texttt{LightweightNeuralNetwork} to \onnx{}.}
    \label{TAB_LWTNN}
    \centering
    \begin{tabular}{cc}
    	\toprule
    	Supported \lwtnn{} layer types & Supported \lwtnn{} activation functions \\ 
    	\midrule
    	Dense & Relu \\
    	Normalization & Sigmoid \\
            & Softmax \\
            & Elu \\
            & Tanh \\
    	\bottomrule
    \end{tabular}
\end{table}

\subsubsection{Running}

A sufficient example for most use-cases is:
\begin{minted}{bash}
  $ petrify-lightweightnn-to-onnx my-lwtnn-file.json -n my-onnx-file 
\end{minted}
Further information on all optional arguments can be accessed via the \texttt{-h} option; most are self-explanatory, but some deserve some additional notes:
\begin{itemize}
	\item \texttt{--float-type} (options: \texttt{float, float16, double}, default: \texttt{float}). This changes the data-type that the \onnx file expects as input and provides as output. This is particularly relevant to \lwtnn{} conversion as \lwtnn{} uses \texttt{double}'s internally, so if you wish to recreate the original \lwtnn{} score close to or beyond single floating-point precision, this may be an option you wish to specify. However, note that 32-bit \texttt{float}s are such a common default for \onnx\ files that some interfaces to \onnxruntime\ in HEP codes may have \texttt{float} hard-coded.
    
    \item \texttt{--opset} (default: \texttt{None}) and \texttt{--ir-version} (default: \texttt{None}): As described for \mvautils{} in section \ref{SUBSUBSEC_MVAUTILSVERSIONS}, although given the maturity of the \onnx operators used for simple deep neural networks (DNNs), using very recent \texttt{opset}s and intermediate-representation versions is less likely to be important in this case.

    \item \texttt{--write-validation} (default: \texttt{False}): See Section~\ref{SUBSUBSEC_LWTNNVALIDATION}.
\end{itemize}

\subsubsection{Validation}

\label{SUBSUBSEC_LWTNNVALIDATION}

The \texttt{--write-validation} option writes a \cpp{} file that performs inference using both the original version of the \lwtnn\ network and the new \onnx\ version, comparing the result.
It is the responsibility of the user to make sure both \lwtnn{} and \onnxruntime\ (neither of which are requirements to run the converter) are correctly installed and linked.
The input features for the inference are randomly generated, using the normalisation information contained within the \lwtnn{} \texttt{json} file to produce inputs at a ``physically reasonable'' scale.
Inference is performed only one set of input features but repeated calls to \petrifyML\ will result in different randomly generated input features unless the random seed is fixed using the \texttt{--seed} flag.
Certainly, this is not a comprehensive validation, but it is a useful first check, and will provide much of the tiresome boiler-plate for the conscientious but potentially time-strapped user who wishes to write a more comprehensive validation script.

\section{Benchmarking}

Benchmarking of \petrifyML\ model conversion performance and output models was performed on a machine with an Intel\textsuperscript{\tiny\textregistered} Core\textsuperscript{\tiny\texttrademark} i7-14700 CPU (28 cores, maximum clock speed \SI{5.4}{GHz}) and \SI{16}{GB} system memory running \AlmaLinux~9.7.
The software environment used \python~3.12.11, with dependency versions summarised in Table~\ref{tab:pythonModuleVersions}.
For \lwtnn, version 2.14.1 was used.
The suite of models used to benchmark the conversion consists of 2~models obtained from \hepdata~\cite{hepdata.134010.v2/r1,hepdata.156121.v1/r1}, 28~models obtained from SimpleAnalysis~\cite{SimpleAnalysisModelData:2026}, 1,173~models obtained from the Systems and Electro Nuclear Scenarios's "Core Library for Advanced Scenario Simulation"~\cite{sens:2026}, and 18~models conceived for testing \petrifyML~\cite{petrifyMLTestdata:2026}.
This amounts to 5~\lwtnn\ neural networks, 6~\mvautils~\lgbm, 3~\mvautils~\xgboost, 3~\sklearn\ and 28~\tmva\ BDTs, and 1,176~\tmva\ MLPs.
Naturally, the sparse availability of models for benchmarking, in particular for \mvautils\ \xgboost\ and \scikitlearn\ BDTs, limits the accuracy of the results.
Nonetheless, some insights can be gained from the results which are presumably sufficiently representative.

\begin{table}
    \caption{List of \python\ modules and their versions used in the benchmarking.}
    \label{tab:pythonModuleVersions}
    \centering
    \begin{tabular}{lr}
        \toprule
            Module & Version \\
        \midrule
            joblib & 1.4.2\\
            json & 2.0.9\\ 
            numpy & 2.1.3\\
            onnx & 1.16.0\\
            onnxruntime &  1.22.0\\
            pandas & 2.2.3\\
            re & 2.2.1\\ 
            ROOT & 6.36.04\\
            sklearn & 1.5.2\\
            tensorflow & 2.19.0\\
            tf2onnx & 1.16.1\\
            tf\_keras & 2.19.0\\
            uproot & 5.6.0\\
        \bottomrule
    \end{tabular}
\end{table}

\subsection{Conversion}

\begin{figure}
    \centering
	\subfloat[]{%
		\includegraphics[width=0.49\textwidth]{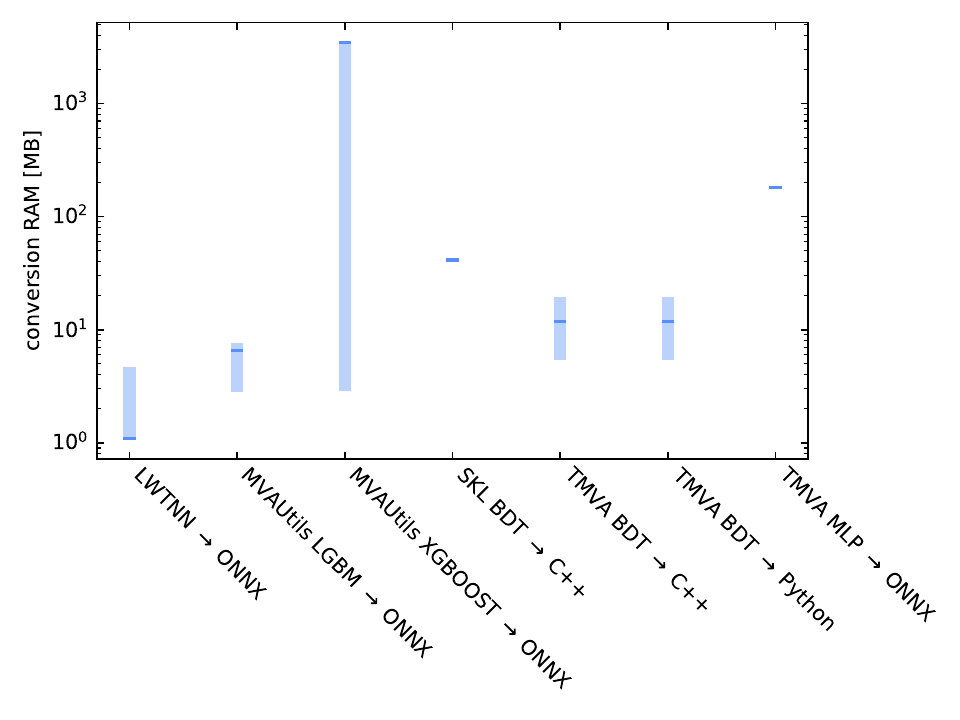}%
	}
	\subfloat[]{%
		\includegraphics[width=0.49\textwidth]{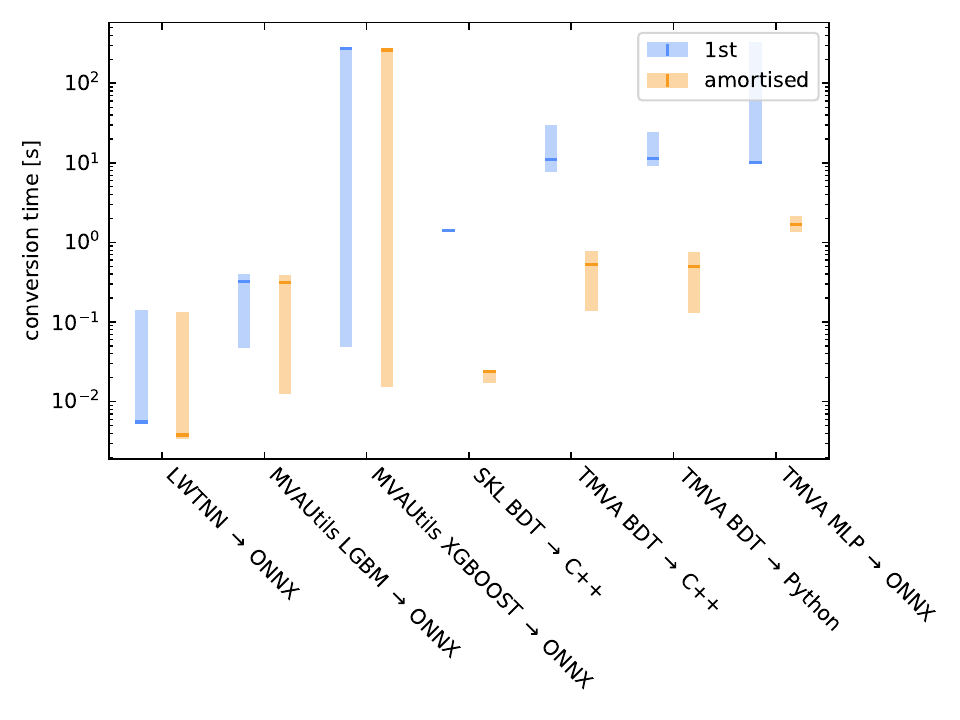}%
	}\\
    \subfloat[]{%
		\includegraphics[width=0.49\textwidth]{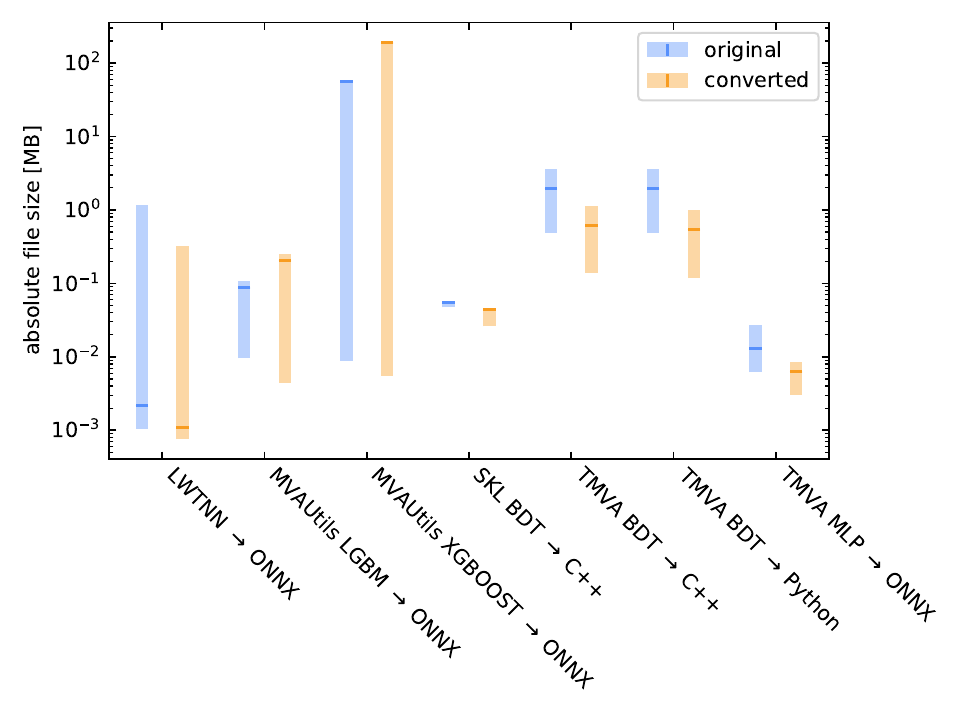}%
	}
    \subfloat[]{%
		\includegraphics[width=0.49\textwidth]{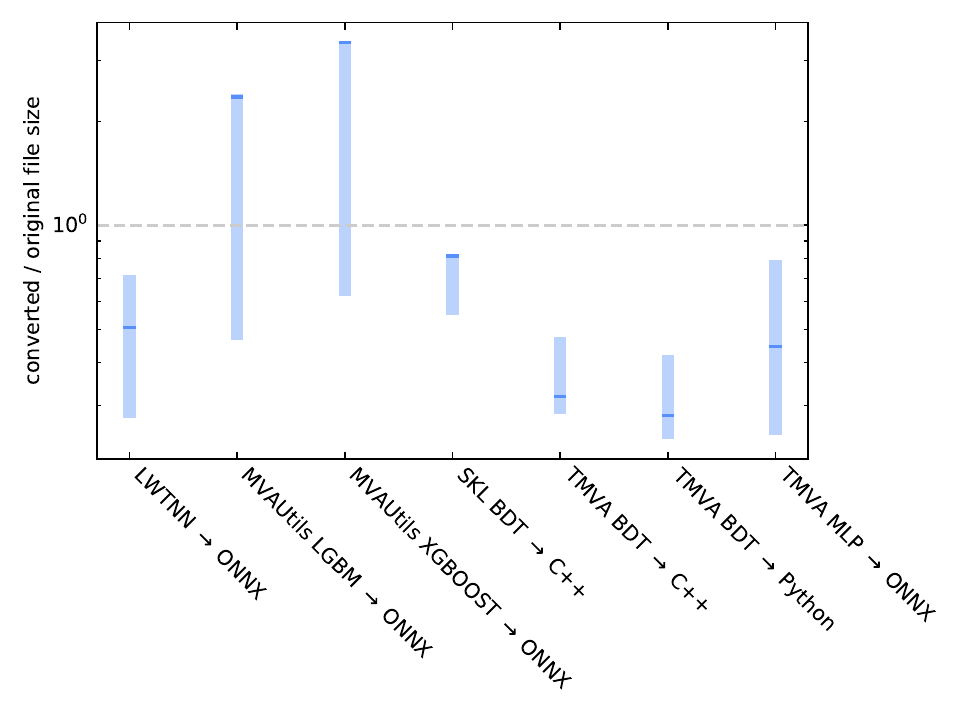}%
	}
    \caption{
        Benchmarking for the different \petrifyML\ conversions.
        (a) Memory usage for the conversion.
        (b) Time for the conversion.
        The time for converting a single model is given in blue, the the amortised time per model for converting 10 models is given in orange.
        (c) Absolute file size of the original file (blue) and converted file (orange).
        (d) Ratio of the converted file size to the original file size.
        The dashed grey line indicates equal file sizes.
        The marker indicates the median observed value, the shaded region the full range from minimum to maximum.
    }
    \label{fig:conversionBenchmark}
\end{figure}

Model conversion using \petrifyML\ was benchmarked with 10 repetitions.
The memory usage for a model conversion in the different channels is given in Figure~\ref{fig:conversionBenchmark}a.
The marker indicates the median observed value, the shaded region the full range from minimum to maximum.
The memory usage extends from a few MB when converting \lwtnn\ neural networks to \onnx\ to about \SI{3.5}{GB} when converting \mvautils\ \xgboost\ BDTs to \onnx.
The latter is presumably an upper limit on the resource usage by \petrifyML\ as well as \mvautils\ \xgboost\ because two out of three tested models consist of forests with the unusually high count of 125,000 BDTs.
As indicated by the sizeable value range, for more regular use cases the memory usage can be smaller than \SI{10}{MB}.
For most conversions with \petrifyML\ to any format, less than \SI{200}{MB} of memory are sufficient.
The small range of observed values for converting \sklearn\ BDTs and \tmva\ MLPs indicates that most of the memory is needed for importing the libraries required for conversion and not for the model conversion itself.
The memory needs do not increase when multiple ML models are converted sequentially.

The conversion time in the different channels is shown in Figure~\ref{fig:conversionBenchmark}b.
The time for converting a single model is given in blue.
The amortised conversion time per model, assessed from 10 successive model conversions, is given in orange.
The lowest values are again observed for converting \lwtnn\ neural networks to \onnx, with a median of about \SI{0.04}{s} for the first conversion.
Similarly, the highest values are again observed for converting \mvautils\ \xgboost\ BDTs to \onnx, with a median of more than \SI{4}{min} for the first conversion.
Notably, successive conversions of \lwtnn\ neural networks or \tmva\ MLPs to \onnx, or \tmva\ BDTs to \cpp\ or \python\ within the same environment can be more than an order of magnitude faster than the first conversion because time-costly module imports only have to be performed once.
For users planning to convert many ML models, it can therefore be advantageous to use \petrifyML's \python\ API for multiple successive conversions instead of performing the conversions one-by-one using the command-line executables.
It is to be noted further, however, that in regular use each conversion only has to be performed once.
Insofar, even sizeable conversion times are presumably of limited concern to the average user.

The absolute original (converted) file sizes in the different channels are shown in Figure~\ref{fig:conversionBenchmark}c in blue (orange).
The marker indicates the median observed value, the shaded region the full range from minimum to maximum.
The lowest values are again observed for \lwtnn\ neural networks, with a median of a few KB.
Similarly, the highest values are again observed for \mvautils\ \xgboost\ BDTs, reaching up to \SI{200}{MB}.
In most cases, only a few MB per model are sufficient.
Converting into native code formats results in about 1,000 lines of \cpp\ code for the tested \sklearn\ BDTs and between 5,000 and 41,000 (4,000 and 31,000) lines of \cpp\ (\python) code for the tested \tmva BDTs.
Interestingly, even these seemingly verbose formats are marginally more space efficient than \sklearn\ BDTs in \texttt{.pkl} or \texttt{.job} format and substantially more space efficient for \tmva\ BDTs in XML format.
The moderate size of all converted files qualifies them well for routine long-term storage of large quantities of ML models.

The ratio of the converted file sizes to the original file sizes in the different channels are shown in Figure~\ref{fig:conversionBenchmark}d.
Overall, the converted files are up to $80\%$ smaller than the original files, apart from \mvautils\ models which are optimised for compactness for use on distributed computing systems.
For these, the converted files are up to a factor 3 larger.

\subsection{Inference}

Model inference in the various formats was benchmarked with 50 repetitions.
No inference benchmarking was performed for \mvautils~models because \mvautils\ is heavily dependent on other ATLAS software packages and running \mvautils\ in standalone mode is close to impossible -- the very shortcoming \petrifyml\ exists to alleviate.
For \tmva~MLP conversions, also the benchmark for the intermediate \keras~models is shown.

\begin{figure}
    \centering
	\subfloat[]{%
		\includegraphics[width=0.49\textwidth]{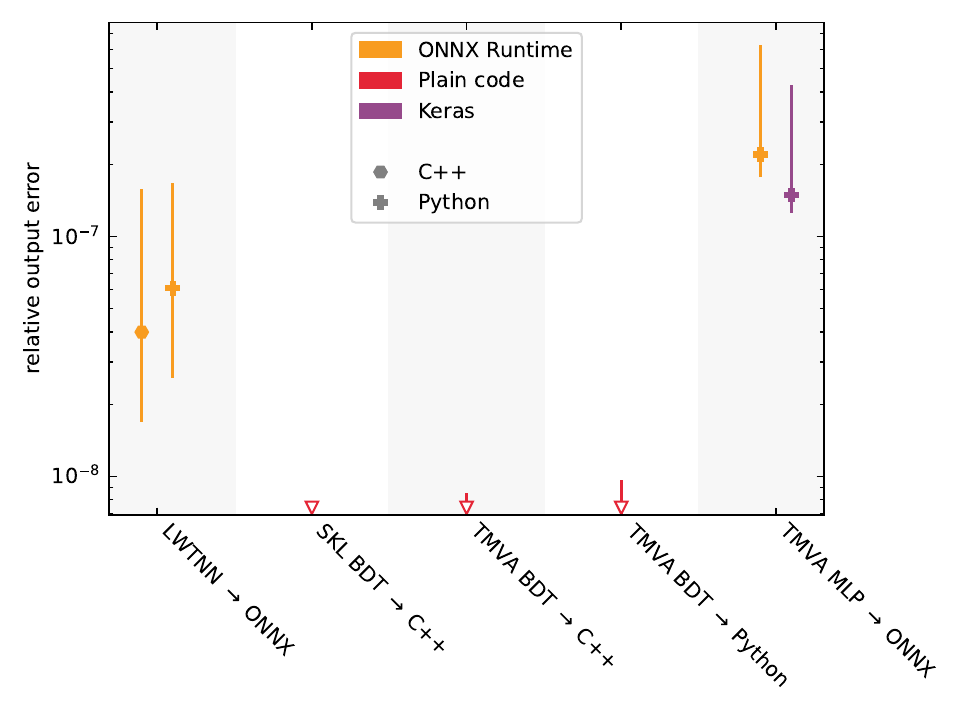}%
	}
	\subfloat[]{%
		\includegraphics[width=0.49\textwidth]{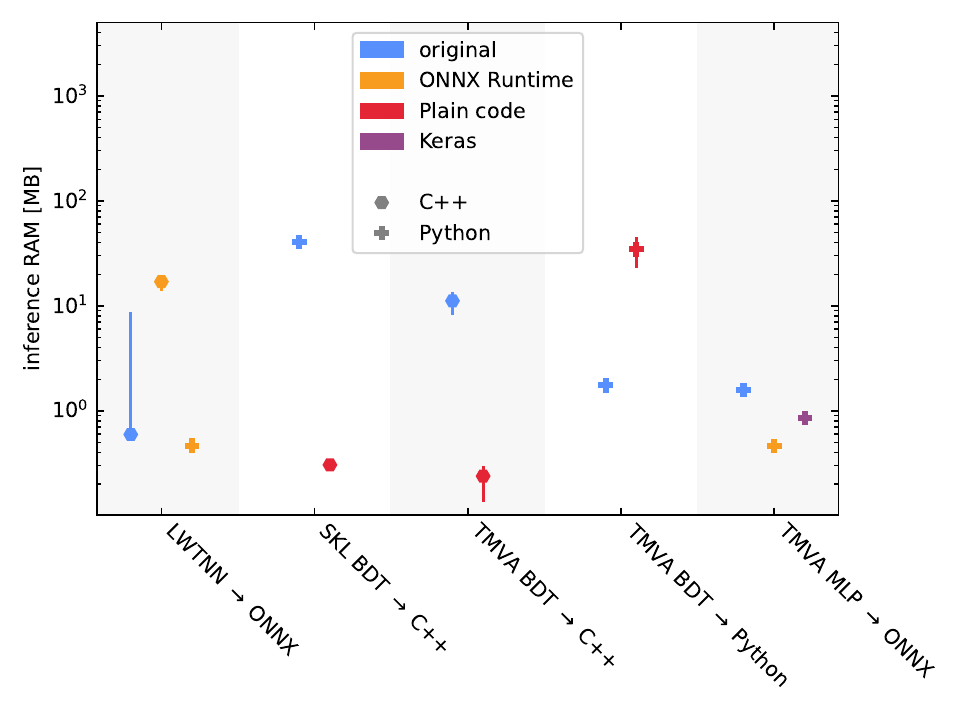}%
	}\\
    \subfloat[]{%
		\includegraphics[width=0.49\textwidth]{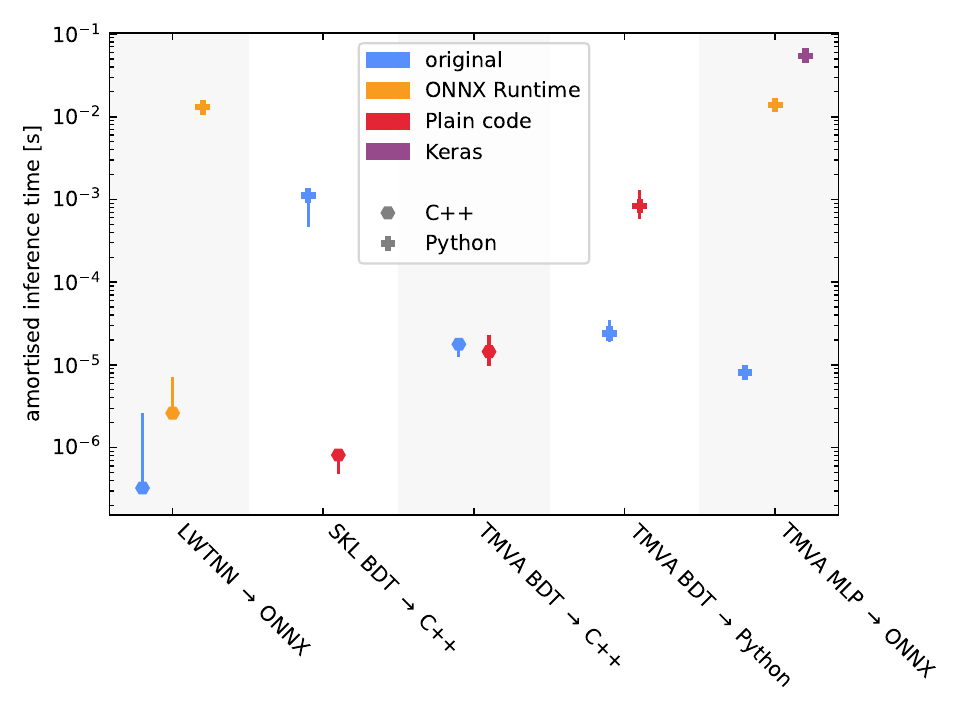}%
	}
    \subfloat[]{%
		\includegraphics[width=0.49\textwidth]{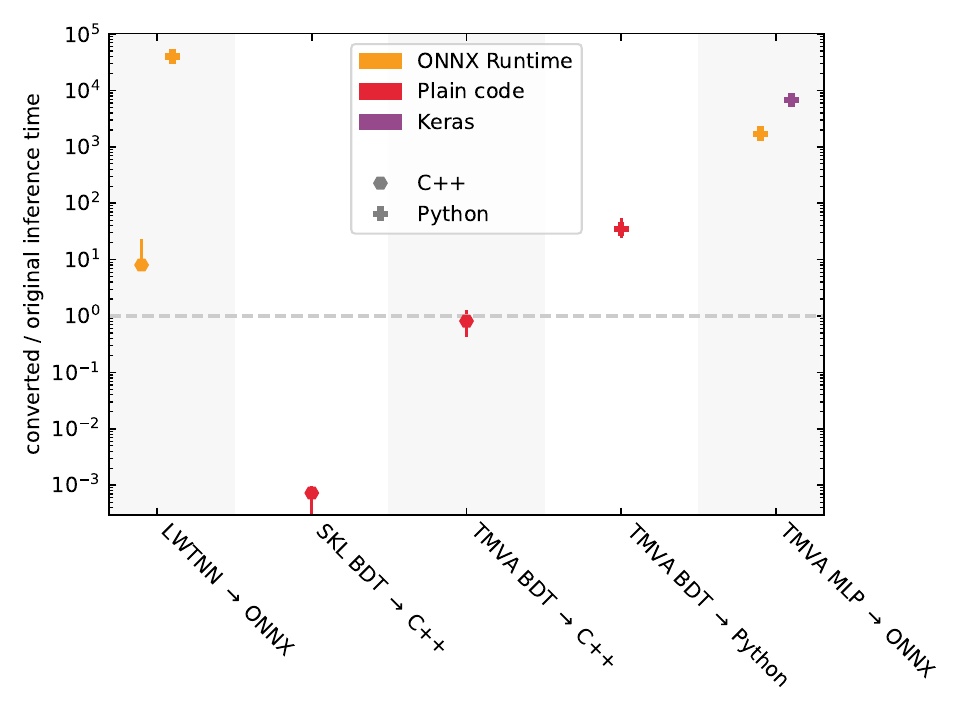}%
	}
    \caption{
        Benchmarking for running inference on the original and converted ML models.
        (a) Relative error of the output of the converted ML~model.
        (b) Memory usage for the inference.
        (c) Amortised inference time.
        (d) Ratio of the inference time of the converted model to the original model.
        The dashed grey line indicates equal inference times.
        Denoted are the median value and the standard deviation.
        The marker colour indicates the environment used for running the model, the marker shape the programming language.
    }
    \label{fig:inferenceBenchmark}
\end{figure}

Figure~\ref{fig:inferenceBenchmark}a gives the relative error of the output of the converted ML~model, determined as the ratio of the difference between the output of the original and converted ML~model to the output of the original ML model.
The output of the converted ML~models deviates from the output of the original model by less than $10^{-6}$ for \onnx\ conversions and agrees perfectly for conversions to plain code.

Figure~\ref{fig:inferenceBenchmark}b shows the memory used for the inference.
In general, less than \SI{100}{MB} are needed for any model.
Particularly noteworthy is that BDTs converted into native \cpp~code are significantly more memory efficient than their original or \python~code counterparts.
The memory needs do not increase when multiple inferences are performed sequentially.

Figure~\ref{fig:inferenceBenchmark}c shows the amortised time for performing a single inference, obtained by averaging over 50~inferences.
In general, less than \SI{0.1}{s} is needed per inference, with values as low as \SI{1}{\micro\second} observed for \lwtnn~models.
Inference in \python\ takes significantly longer than inference in \cpp, showcasing the advantage of compiled code.
The exception is \tmva\ were inference in \python\ and \cpp\ is similarly fast, presumably because a compiled backend is used independent of the frontend language.
No significant difference between performing the first and subsequent inferences was observed.

Figure~\ref{fig:inferenceBenchmark}d shows the relative inference time, calculated as the ratio between performing inference on the converted model to performing inference on the original model.
Performing inference using the converted model is generally slower, particularly in \python.
The main exception is plain \cpp~code for which performing BDT inference can be as fast (\tmva) or significantly faster (\sklearn) than using the original model.
However, given the absolute inference time is still small in any case, the relative gains or losses when using the converted models are presumably negligible.

\section{Summary}

We have introduced the \petrifyml package for converting configurations from commonly used HEP ML tools to either native \python or \cpp\ code, or the industry-standard \onnx format.
The package has few and lightweight dependencies.
In combination with the easy installation via \pip\ and straightforward command-line interfaces, this makes it well suited to simplify the long-term preservation of HEP ML models.

At the time of writing, the package can convert all non-standard model formats published by the \atlas\ collaboration without native \onnx\ support\footnote{We are not aware of any pertinent examples from \cms\ that we could naturally support, but we would like to emphasise that we are equally keen to ensure the long-term re-usability of ML models from across the \lhc\ or even outside HEP.}.
Future developments might extend the functionality to further ML models, \eg from \lwtnn{} or \mvautils{}, if the need arises.

\section{Acknowledgements}

AB, MH and TP acknowledge funding from OpenMAPP project, via UKRI/EPSRC EP/Y036360/1 and
National Science Centre, Poland under CHIST-ERA programme (NCN 2022/04/Y/ST2/00186). TP gratefully acknowledges Polish high-performance computing infrastructure PLGrid (HPC Center: ACK Cyfronet AGH) for providing computer facilities and support within computational grants no. PLG/2025/018203 and PLG/2026/019538.

\printbibliography

@article{Araz:2023mda,
    author = "Araz, Jack Y. and others",
    title = "{Les Houches guide to reusable ML models in LHC analyses}",
    eprint = "2312.14575",
    archivePrefix = "arXiv",
    primaryClass = "hep-ph",
    doi = "10.21468/SciPostPhysCommRep.3",
    month = "12",
    year = "2023"
}

@software{daniel_hay_guest_2024_14276439,
  author       = {Daniel Hay Guest and
                  Joshua Wyatt Smith and
                  Michela Paganini and
                  Michael Kagan and
                  Marie Lanfermann and
                  Attila Krasznahorkay and
                  Daniel Edison Marley and
                  Aishik Ghosh and
                  Benjamin Huth and
                  Matthew Feickert},
  title        = "{lwtnn/lwtnn: v2.14.1}",
  month        = dec,
  year         = 2024,
  publisher    = {Zenodo},
  version      = {v2.14.1},
  doi          = {10.5281/zenodo.14276439},
  url          = {https://doi.org/10.5281/zenodo.14276439},
  swhid        = {swh:1:dir:e8e8cd6fad6becd5e8b193c1148f94e023df3bae
                   ;origin=https://doi.org/10.5281/zenodo.597221;visi
                   t=swh:1:snp:fe7e271159192d7ee02e85798322462a200620
                   a1;anchor=swh:1:rel:698f80ceb7827b6d852a1bbf695eaf
                   b0fccc5079;path=/
                  },
}

@article{ATLAS:2024tqe,
    author = "Aad, Georges and others",
    collaboration = "ATLAS",
    title = "{Search for pair production of higgsinos in events with two Higgs bosons and missing transverse momentum in s=13{\,}{\,}TeV pp collisions at the ATLAS experiment}",
    eprint = "2401.14922",
    archivePrefix = "arXiv",
    primaryClass = "hep-ex",
    reportNumber = "CERN-EP-2023-278",
    doi = "10.1103/PhysRevD.109.112011",
    journal = "Phys. Rev. D",
    volume = "109",
    number = "11",
    pages = "112011",
    year = "2024"
}

@misc{hepdata.136030.v1,
    author = "Aad, Georges and others",
    collaboration = "{ATLAS}",
    title = "{HEPData entry for 'Search for pair production of higgsinos in events with two Higgs bosons and missing transverse momentum in $\sqrt{s}=13$ TeV $pp$ collisions at the ATLAS experiment' (Version 1)}",
    howpublished = "{HEPData (collection)}",
    year = 2024,
    note = "\url{https://doi.org/10.17182/hepdata.136030.v1}"
}

@article{ATLAS:2022hnn,
    author = "Aad, Georges and others",
    collaboration = "ATLAS",
    title = "{Search for pair-production of vector-like quarks in pp collision events at $\sqrt{s}=13$ TeV with at least one leptonically decaying Z boson and a third-generation quark with the ATLAS detector}",
    eprint = "2210.15413",
    archivePrefix = "arXiv",
    primaryClass = "hep-ex",
    reportNumber = "CERN-EP-2021-207",
    doi = "10.1016/j.physletb.2023.138019",
    journal = "Phys. Lett. B",
    volume = "843",
    pages = "138019",
    year = "2023"
}

@misc{hepdata.134010.v1,
    author = "Aad, Georges and others",
    collaboration = "ATLAS",
    title = "{HEPData entry for 'Search for pair-production of vector-like quarks in $pp$ collision events at $\sqrt{s}=13$ TeV with at least one leptonically decaying $Z$ boson and a third-generation quark with the ATLAS detector' (Version 1)}",
    howpublished = "{HEPData (collection)}",
    year = 2025,
    note = "\url{https://doi.org/10.17182/hepdata.134010.v1}"
}

@article{ATLAS:2019bwq,
    author = "Aad, Georges and others",
    collaboration = "ATLAS",
    title = "{ATLAS b-jet identification performance and efficiency measurement with $t{\bar{t}}$ events in pp collisions at $\sqrt{s}=13$ TeV}",
    eprint = "1907.05120",
    archivePrefix = "arXiv",
    primaryClass = "hep-ex",
    reportNumber = "CERN-EP-2019-132",
    doi = "10.1140/epjc/s10052-019-7450-8",
    journal = "Eur. Phys. J. C",
    volume = "79",
    number = "11",
    pages = "970",
    year = "2019"
}

@article{CMS:2020poo,
    author = "Sirunyan, Albert M and others",
    collaboration = "CMS",
    title = "{Identification of heavy, energetic, hadronically decaying particles using machine-learning techniques}",
    eprint = "2004.08262",
    archivePrefix = "arXiv",
    primaryClass = "hep-ex",
    reportNumber = "CMS-JME-18-002, CERN-EP-2020-037",
    doi = "10.1088/1748-0221/15/06/P06005",
    journal = "JINST",
    volume = "15",
    number = "06",
    pages = "P06005",
    year = "2020"
}

@article{Quinlan:1986gsq,
    author = "Quinlan, J. R.",
    title = "{Induction of decision trees}",
    doi = "10.1007/BF00116251",
    journal = "Machine Learning",
    volume = "1",
    number = "1",
    pages = "81--106",
    year = "1986"
}

@article{Corneliusen:1989zw,
    author = "Corneliusen, A. and Terdal, P. and Knight, T. and Spencer, J.",
    title = "{COMPUTATION AND CONTROL WITH NEURAL NETS}",
    reportNumber = "SLAC-PUB-5035",
    doi = "10.1016/0168-9002(90)91491-S",
    journal = "Nucl. Instrum. Meth. A",
    volume = "293",
    pages = "507--516",
    year = "1990"
}

@article{Speckmayer:2010zz,
    author = "Speckmayer, P. and Hocker, A. and Stelzer, J. and Voss, H.",
    editor = "Gruntorad, Jan and Lokajicek, Milos",
    title = "{The toolkit for multivariate data analysis, TMVA 4}",
    doi = "10.1088/1742-6596/219/3/032057",
    journal = "J. Phys. Conf. Ser.",
    volume = "219",
    pages = "032057",
    year = "2010"
}

@article{Dercks:2016npn,
    author = "Dercks, Daniel and Desai, Nishita and Kim, Jong Soo and Rolbiecki, Krzysztof and Tattersall, Jamie and Weber, Torsten",
    title = "{CheckMATE 2: From the model to the limit}",
    eprint = "1611.09856",
    archivePrefix = "arXiv",
    primaryClass = "hep-ph",
    reportNumber = "CTPU-16-36, CSIC-16-116, TTK-16-47",
    doi = "10.1016/j.cpc.2017.08.021",
    journal = "Comput. Phys. Commun.",
    volume = "221",
    pages = "383--418",
    year = "2017"
}

@software{ATLAS:2019ath,
    author = "Aad, Georges and others",
    collaboration = "ATLAS",
    title = "{Athena}",
    doi = "10.5281/zenodo.2641997",
    year = 2019
}

@article{Brun:1997pa,
    author = "Brun, R. and Rademakers, F.",
    editor = "Werlen, M. and Perret-Gallix, D.",
    title = "{ROOT: An object oriented data analysis framework}",
    doi = "10.1016/S0168-9002(97)00048-X",
    journal = "Nucl. Instrum. Meth. A",
    volume = "389",
    pages = "81--86",
    year = "1997"
}

@misc{chollet2015keras,
  title = "{Keras}",
  author = {Chollet, Fran\c{c}ois and others},
  year={2015},
  url = "https://keras.io",
}

@misc{onnxruntime,
  title = "{ONNX Runtime}",
  author="{ONNX Runtime developers}",
  year={2021},
  url = "https://onnxruntime.ai",
  note={Version: 1.18.0}
}

@misc{tensorflow2015-whitepaper,
    title = "{TensorFlow: Large-Scale Machine Learning on Heterogeneous Systems}",
    url={https://www.tensorflow.org/},
    note={Software available from tensorflow.org},
    author={
        Mart\'{i}n~Abadi and
        Ashish~Agarwal and
        Paul~Barham and
        Eugene~Brevdo and
        Zhifeng~Chen and
        Craig~Citro and
        Greg~S.~Corrado and
        Andy~Davis and
        Jeffrey~Dean and
        Matthieu~Devin and
        Sanjay~Ghemawat and
        Ian~Goodfellow and
        Andrew~Harp and
        Geoffrey~Irving and
        Michael~Isard and
        Yangqing Jia and
        Rafal~Jozefowicz and
        Lukasz~Kaiser and
        Manjunath~Kudlur and
        Josh~Levenberg and
        Dandelion~Man\'{e} and
        Rajat~Monga and
        Sherry~Moore and
        Derek~Murray and
        Chris~Olah and
        Mike~Schuster and
        Jonathon~Shlens and
        Benoit~Steiner and
        Ilya~Sutskever and
        Kunal~Talwar and
        Paul~Tucker and
        Vincent~Vanhoucke and
        Vijay~Vasudevan and
        Fernanda~Vi\'{e}gas and
        Oriol~Vinyals and
        Pete~Warden and
        Martin~Wattenberg and
        Martin~Wicke and
        Yuan~Yu and
        Xiaoqiang~Zheng},
  year={2015},
}

@software{tf2onnx,
    author = "{tf2onnx developers}",
    title = "{tf2onnx - Convert TensorFlow, Keras, Tensorflow.js and Tflite models to ONNX}",
    url = {https://github.com/onnx/tensorflow-onnx}
}

@article{scikit-learn,
  title = "{Scikit-learn: Machine Learning in Python}",
  author={Pedregosa, F. and Varoquaux, G. and Gramfort, A. and Michel, V.
          and Thirion, B. and Grisel, O. and Blondel, M. and Prettenhofer, P.
          and Weiss, R. and Dubourg, V. and Vanderplas, J. and Passos, A. and
          Cournapeau, D. and Brucher, M. and Perrot, M. and Duchesnay, E.},
  journal={Journal of Machine Learning Research},
  volume={12},
  pages={2825--2830},
  year={2011}
}

@Article{harris2020array,
 title         = "{Array programming with NumPy}",
 author        = {Charles R. Harris and K. Jarrod Millman and St{\'{e}}fan J.
                 van der Walt and Ralf Gommers and Pauli Virtanen and David
                 Cournapeau and Eric Wieser and Julian Taylor and Sebastian
                 Berg and Nathaniel J. Smith and Robert Kern and Matti Picus
                 and Stephan Hoyer and Marten H. van Kerkwijk and Matthew
                 Brett and Allan Haldane and Jaime Fern{\'{a}}ndez del
                 R{\'{i}}o and Mark Wiebe and Pearu Peterson and Pierre
                 G{\'{e}}rard-Marchant and Kevin Sheppard and Tyler Reddy and
                 Warren Weckesser and Hameer Abbasi and Christoph Gohlke and
                 Travis E. Oliphant},
 year          = {2020},
 month         = sep,
 journal       = {Nature},
 volume        = {585},
 number        = {7825},
 pages         = {357--362},
 doi           = {10.1038/s41586-020-2649-2},
 publisher     = {Springer Science and Business Media {LLC}},
 url           = {https://doi.org/10.1038/s41586-020-2649-2}
}

@software{reback2020pandas,
    author       = "{pandas development team}",
    title        = "{pandas}",
    month        = feb,
    year         = 2020,
    publisher    = {Zenodo},
    version      = {latest},
    doi          = {10.5281/zenodo.3509134},
    url          = {https://doi.org/10.5281/zenodo.3509134}
}

@software{uproot2020,
    author = {Jim Pivarski},
    title = "{scikit-hep/uproot: 3.12.0}",
    publisher = {Zenodo},
    year = 2020,
    month = 7,
    day = 20,
    doi = {10.5281/zenodo.3952728}
}

@software{pytest,
    title        = "{pytest}",
    author       = {Holger Krekel and Bruno Oliveira and Ronny Pfannschmidt and Floris Bruynooghe and Brianna Laugher and Florian Bruhin},
    year         = {2004},
    url          = {https://github.com/pytest-dev/pytest},
}

@inproceedings{NIPS2017_6449f44a,
    author = {Ke, Guolin and Meng, Qi and Finley, Thomas and Wang, Taifeng and Chen, Wei and Ma, Weidong and Ye, Qiwei and Liu, Tie-Yan},
    booktitle = {Advances in Neural Information Processing Systems},
    editor = {I. Guyon and U. Von Luxburg and S. Bengio and H. Wallach and R. Fergus and S. Vishwanathan and R. Garnett},
    pages = {},
    publisher = {Curran Associates, Inc.},
    title = "{LightGBM: A Highly Efficient Gradient Boosting Decision Tree}",
    url = {https://proceedings.neurips.cc/paper_files/paper/2017/file/6449f44a102fde848669bdd9eb6b76fa-Paper.pdf},
    volume = {30},
    year = {2017}
}

@inproceedings{Chen:2016:XST:2939672.2939785,
    author = {Chen, Tianqi and Guestrin, Carlos},
    title = "{XGBoost: A Scalable Tree Boosting System}",
    booktitle = {Proceedings of the 22nd ACM SIGKDD International Conference on Knowledge Discovery and Data Mining},
    series = {KDD '16},
    year = {2016},
    isbn = {978-1-4503-4232-2},
    location = {San Francisco, California, USA},
    pages = {785--794},
    numpages = {10},
    url = {http://doi.acm.org/10.1145/2939672.2939785},
    doi = {10.1145/2939672.2939785},
    acmid = {2939785},
    publisher = {ACM},
    address = {New York, NY, USA},
}

@misc{onnx_github,
    author = "{ONNX development team}",
    title = "{onnx: Open Neural Network Exchange}",
    url = "https://github.com/onnx/onnx",
    note = {Accessed: 2025-09-01},
}

@article{ATLAS:2022yru,
    author="Aad et al.",
    collaboration = "ATLAS",
    title = "{SimpleAnalysis: Truth-level Analysis Framework}",
    reportNumber = "ATL-PHYS-PUB-2022-017",
    doi = "10.17181/CERN.R6S3.0QKV",
    year = "2022"
}

@article{ATLAS:2024woy,
    author = "Aad, Georges and others",
    collaboration = "ATLAS",
    title = "{Search for supersymmetry using vector boson fusion signatures and missing transverse momentum in pp collisions at $ \sqrt{s} $ = 13 TeV with the ATLAS detector}",
    eprint = "2409.18762",
    archivePrefix = "arXiv",
    primaryClass = "hep-ex",
    reportNumber = "CERN-EP-2024-246",
    doi = "10.1007/JHEP12(2024)116",
    journal = "JHEP",
    volume = "12",
    pages = "116",
    year = "2024"
}

@misc{hepdata.156776.v1,
    author = "{ATLAS Collaboration}",
    title = "{Search for supersymmetry using vector boson fusion signatures and missing transverse momentum in $pp$ collisions at $\sqrt{s}=13$ TeV with the ATLAS detector (Version 1)}",
    howpublished = "{HEPData (collection)}",
    year = 2025,
    note = "\url{https://doi.org/10.17182/hepdata.156776.v1}"
}

@article{Bierlich:2024vqo,
    author = "Bierlich, Christian and Buckley, Andy and Butterworth, Jonathan Mark and Gutschow, Christian and Lonnblad, Leif and Procter, Tomasz and Richardson, Peter and Yeh, Yoran",
    title = "{Robust independent validation of experiment and theory: Rivet version 4 release note}",
    eprint = "2404.15984",
    archivePrefix = "arXiv",
    primaryClass = "hep-ph",
    reportNumber = "MCNET-24-05",
    doi = "10.21468/SciPostPhysCodeb.36",
    journal = "SciPost Phys. Codeb.",
    volume = "36",
    pages = "1",
    year = "2024"
}

@misc{hepdata.134010.v2/r1,
    author = "{ATLAS Collaboration}",
    collaboration = "ATLAS",
    title = "{'MCBOT\_rel21\_wp77.json' of 'Search for pair-production of vector-like quarks in $pp$ collision events at $\sqrt{s}=13$ TeV with at least one leptonically decaying $Z$ boson and a third-generation quark with the ATLAS detector' (Version 2)}",
    howpublished = "{HEPData (other)}",
    year = 2025,
    note = "\url{https://doi.org/10.17182/hepdata.134010.v2/r1}"
}

@misc{hepdata.156121.v1/r1,
    author = "{CMS Collaboration}",
    collaboration = "CMS",
    title = "{'nano\_train\_model.json' of 'General search for supersymmetric particles in scenarios with compressed mass spectra using proton-proton collisions at $\sqrt{s}$ = 13 TeV' (Version 1)}",
    howpublished = "{HEPData (other)}",
    year = 2025,
    note = "\url{https://doi.org/10.17182/hepdata.156121.v1/r1}"
}

@misc{SimpleAnalysisModelData:2026,
    author="Aad et al.",
    collaboration = "ATLAS",
    title = "{SimpleAnalysis: Truth-level Analysis Framework -- data}",
    year = "2026",
    url = "https://gitlab.cern.ch/atlas-sa/simple-analysis/-/tree/aa8efaa8b73b0df6d1ac9089ff6acdb386aeabe0/SimpleAnalysisCodes/data"
}

@misc{petrifyMLTestdata:2026,
    author = "A. Buckley and L. Corpe and M. Habedank and T. Procter",
    title = "{petrifyML testdata}",
    year = 2026,
    url = "https://gitlab.com/hepcedar/petrifyML/-/tree/4572b045ffcb0b16c2bb04eb70ede7675fc83125/tests/testdata"
}

@misc{sens:2026,
    author = "{Centre National de la Recherche Scientifique, Institut National de Physique Nucléaire et de Physique des Particules, Institut de Radioprotection et de Sûreté Nucléaire}",
    title = "Systems and Electro Nuclear Scenarios: Core Library for Advanced Scenario Simulation",
    year = 2026,
    url = "https://gitlab-preprod.in2p3.fr/sens/CLASS/-/tree/972c2db641181549bc381efa4d7752fac24dc12a/DATA_BASES"
}

@software{The_joblib_developers_joblib,
    author = "{joblib developers}",
    doi = {https://doi.org/10.5281/zenodo.14915601},
    title = {{joblib}},
    url = {https://github.com/joblib/joblib},
    version = {latest}
}

@article{Alwall:2014hca,
    author = "Alwall, J. and Frederix, R. and Frixione, S. and Hirschi, V. and Maltoni, F. and Mattelaer, O. and Shao, H. -S. and Stelzer, T. and Torrielli, P. and Zaro, M.",
    title = "{The automated computation of tree-level and next-to-leading order differential cross sections, and their matching to parton shower simulations}",
    eprint = "1405.0301",
    archivePrefix = "arXiv",
    primaryClass = "hep-ph",
    reportNumber = "CERN-PH-TH-2014-064, CP3-14-18, LPN14-066, MCNET-14-09, ZU-TH-14-14",
    doi = "10.1007/JHEP07(2014)079",
    journal = "JHEP",
    volume = "07",
    pages = "079",
    year = "2014"
}

@article{Frederix:2018nkq,
    author = "Frederix, R. and Frixione, S. and Hirschi, V. and Pagani, D. and Shao, H. -S. and Zaro, M.",
    title = "{The automation of next-to-leading order electroweak calculations}",
    eprint = "1804.10017",
    archivePrefix = "arXiv",
    primaryClass = "hep-ph",
    reportNumber = "Nikhef/2018-015, TUM-HEP-1138/18, NIKHEF-2018-015, TUM-HEP-1138-18",
    doi = "10.1007/JHEP11(2021)085",
    journal = "JHEP",
    volume = "07",
    pages = "185",
    year = "2018",
    note = "[Erratum: JHEP 11, 085 (2021)]"
}

@booklet{protos,
	title = "{PROTOS, a PROgram for TOp Simulations}",
	author = "{J. A. Aguilar-Saavedra}",
	note="\url{http://jaguilar.web.cern.ch/jaguilar/protos/}",
}

@misc{emlearn,
  author       = {Nordby, Jon AND Cooke, Mark AND Horvath, Adam},
  title        = "{emlearn: Machine Learning inference engine for Microcontrollers and Embedded Devices}",
  month        = mar,
  year         = 2019,
  doi          = {10.5281/zenodo.2589394},
  url          = {https://doi.org/10.5281/zenodo.2589394}
}

@software{fastml_hls4ml,
  author       = "{FastML Team}",
  title        = {fastmachinelearning/hls4ml},
  year         = 2025,
  publisher    = {Zenodo},
  version      = {v1.2.0},
  doi          = {10.5281/zenodo.1201549},
  url          = {https://github.com/fastmachinelearning/hls4ml}
}

@article{Duarte:2018ite,
    author = "Duarte, Javier and others",
    title = "{Fast inference of deep neural networks in FPGAs for particle physics}",
    eprint = "1804.06913",
    archivePrefix = "arXiv",
    primaryClass = "physics.ins-det",
    reportNumber = "FERMILAB-PUB-18-089-E",
    doi = "10.1088/1748-0221/13/07/P07027",
    journal = "JINST",
    volume = "13",
    number = "07",
    pages = "P07027",
    year = "2018"
}

@software{onnx2c,
  author = {Kraiskil},
  title = "{onnx2c: A compiler to convert ONNX weights and model to static C code}",
  year = {2021},
  publisher = {GitHub},
  journal = {GitHub repository},
  url = "https://github.com/kraiskil/onnx2c",
  commit = {master}
}

@software{petrifyml,
  author = {A. Buckley and L. Corpe and M. Habedank and T. Procter},
  title = "{petrifyML v2.1.0}",
  year = {2026},
  publisher = {GitLab},
  journal = {GitLab repository},
  doi = "https://doi.org/10.25323/openmapp.petrifyml.v2.1.0",
}

@article{Bierlich:2019rhm,
    author = "Bierlich, Christian and others",
    title = "{Robust Independent Validation of Experiment and Theory: Rivet version 3}",
    eprint = "1912.05451",
    archivePrefix = "arXiv",
    primaryClass = "hep-ph",
    reportNumber = "MCnet-19-26",
    doi = "10.21468/SciPostPhys.8.2.026",
    journal = "SciPost Phys.",
    volume = "8",
    pages = "026",
    year = "2020"
}

@article{GAMBIT:2017qxg,
    author = "Bal{\'a}zs, Csaba and others",
    collaboration = "GAMBIT",
    title = "{ColliderBit: a GAMBIT module for the calculation of high-energy collider observables and likelihoods}",
    eprint = "1705.07919",
    archivePrefix = "arXiv",
    primaryClass = "hep-ph",
    reportNumber = "gambit-code-2017",
    doi = "10.1140/epjc/s10052-017-5285-8",
    journal = "Eur. Phys. J. C",
    volume = "77",
    number = "11",
    pages = "795",
    year = "2017"
}

@article{GAMBIT:2017yxo,
    author = "Athron, Peter and others",
    collaboration = "GAMBIT",
    title = "{GAMBIT: The Global and Modular Beyond-the-Standard-Model Inference Tool}",
    eprint = "1705.07908",
    archivePrefix = "arXiv",
    primaryClass = "hep-ph",
    reportNumber = "COEPP-MN-17-6, NORDITA-2017-074, DESY-17-236, CERN-TH-2017-166, CoEPP-MN-17-6, NORDITA 2017-074, gambit-code-2017",
    doi = "10.1140/epjc/s10052-017-5321-8",
    journal = "Eur. Phys. J. C",
    volume = "77",
    number = "11",
    pages = "784",
    year = "2017",
    note = "[Addendum: Eur.Phys.J.C 78, 98 (2018)]"
}

\newpage
\begin{appendices}

\section{A petrify-lightweightnn-to-onnx example: ATLAS-EXOT-2018-58 in \rivet{}}

Reference~\cite{ATLAS:2022hnn} is an ATLAS search for pair-produced vector-like quarks decaying to a $Z$-boson and at least one third-generation quark. Essential to the analysis was the tagger ``MCBOT'' (Multi-Class Boosted Object Tagger), a DNN which was used to classify large radius jets as originating from a Higgs boson, vector boson, top quark, or else as a ``light jet''. After an initial event selection established a broader signal region, the combinations of jet tags present was used to subdivide the signal region into seven (five) sub-regions in the two-lepton (three-lepton) channel. Kinematic distributions in these sub-regions where then used as the final discriminants in the analysis's fit.

In this case, it is effectively impossible to reinterpret the analysis without access to the neural network weights.
While efficiency figures are given for MCBOT, because the true efficiencies are not uniform in the jet kinematics (as demonstrated by Figure~3 in Reference~\cite{ATLAS:2022hnn}), randomly sampling using these efficiencies would bias the kinematic distributions used in the final fit beyond usability.

\begin{figure}[htb]
    \centering
    \includegraphics[width=0.5\columnwidth]{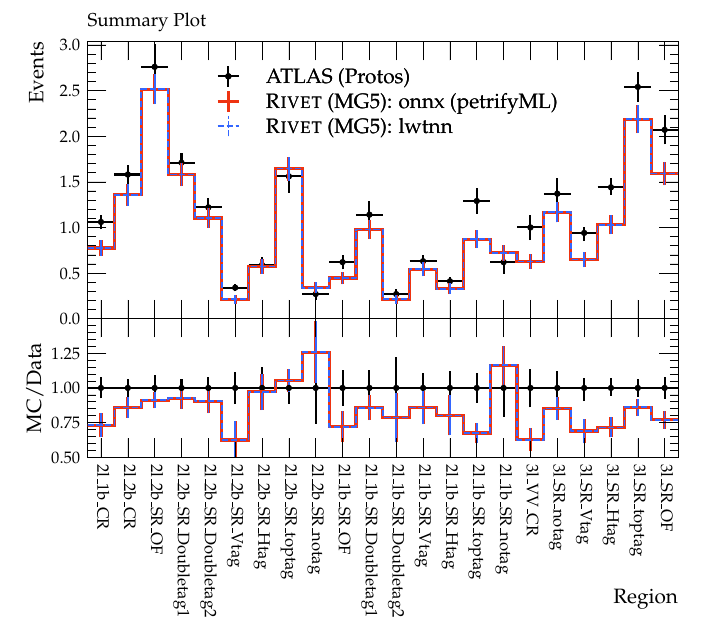}
    \caption{Comparison of the original \lwtnn{} network and its petrified \onnx{} form inside the \rivet{} implementation of Reference~\cite{ATLAS:2022hnn}. Both \rivet{} analyses were run over the same 190k signal events, generated with \madgraph~\cite{Alwall:2014hca,Frederix:2018nkq}. The ATLAS prediction for the same signal (singlet-$T\bar{T}$ pair production), obtained from \hepdata and originally generated with \protos~\cite{protos} is displayed for comparison.}
    \label{fig:LWTNN_RIVET_EXAMPLE}
\end{figure}

The DNN was provided by ATLAS in the form of an \lwtnn{} file on \hepdata~\cite{hepdata.134010.v1}, and converted to \onnx{} format using \petrifyML{}.
In order to validate the implementation in \onnx{} used in the \rivet{} analysis, a version of the analysis which called \lwtnn{} directly (using the now deprecated \rivet{}-\lwtnn{} interface) was also written to serve as a direct comparison. The results, shown in Figure~\ref{fig:LWTNN_RIVET_EXAMPLE}, show exact event-for-event replication\footnote{Note that this level of precision required the \onnx{} float type to be set to double precision, to match the output type of \lwtnn{}.}. Timing differences between the tools were sufficiently small that they could not be measured: within the \rivet{} event loop, tasks such as jet-clustering will almost always be more expensive than ML inference.

\end{appendices}

\end{document}